# Managing Uncertainty: A Case for Probabilistic Grid Scheduling

Aleksandar Lazarević, Lionel Sacks, Ognjen Prnjat

Dept. of Electronic and Electrical Engineering, University College London, Torrington Place, London WC1E 7JE, UK
{a.lazarevic, lsacks, oprnjat}@ee.ucl.ac.uk

**Abstract.** The Grid technology is evolving into a global, service-orientated architecture – a universal platform for delivering future high demand computational services. Strong adoption of the Grid and the utility computing concept is leading to an increasing number of Grid installations running a wide range of applications of different size and complexity. In this paper we address the problem of delivering deadline/economy based scheduling in a heterogeneous application environment using statistical properties of job's historical executions and its associated meta-data. This approach is motivated by a study of six-month computational load generated by Grid applications in a multi-purpose Grid cluster serving a community of twenty e-Science projects. The observed job statistics, resource utilisation and user behaviour is discussed in the context of management approaches and models most suitable for supporting a probabilistic and autonomous scheduling architecture.

## 1. Introduction

Increasing demand for high-performance computer systems in recent years has helped establish Grid technology as an attractive choice for academic research clusters and high-demand business computing alike [1]. But proliferation of Grid installations, encouraged by low entry barriers typical of off-the-shelf components and open source licensing, has also meant that Grid is evolving into a global, service-orientated utility platform. Grid's migration from cutting edge to the mainstream is already evident, and is driven in part by the financial benefits (such as reduced total cost of ownership and better value through economies of scale) as well as policy shifts within large computational consumers. As this shift takes place, an associated change of the Grid application landscape will follow. Our work is focused on enabling a more efficient and user-friendlier scheduling in a future Grid serving a rich mix of diverse e-Science applications. We will argue that in such environments, a combination of deadline and economy based scheduling can be delivered by utilising statistical models of application execution times and resource requirements based on historical data, meta-data normally associated with submitted jobs, and the current state of the Grid system.

The need for a more efficient scheduling has been discussed previously [2-5]. Departing from a batch model inherited from legacy cluster systems, a deadline/economy based system more suited to human workflow would allow users to specify a deadline and a nominative price by which they would expect their job finished. Such scheduling systems could not be built unless certain predictions could be made on the length of execution and resource requirements of the jobs pending in the queue. Previous research in predicting the execution times has focused on deep analysis of the application source code [6], instrumentation of the application, or re-linking to specialised prediction libraries customised for both the type of the application and its target hardware [7]. Some of these methods have given encouraging results [8], but the development, deployment and administration effort can only be justified in cases of highest-end computational resources.

We have taken the view that in delivering predictions in a varied application landscape executing on a utility Grid platform, spot prediction accuracy would come second to the prediction speed and the level of administrator intervention. With this in mind, we have strived to develop a significantly self-managing system. Working in a heterogeneous application space, and delivering statistics-based predictions requires a good understanding of computational load presented to a Grid cluster. Studies of production, multi-purpose Grids are rare, and the interaction and multiplexing effects of various applications are unknown. Therefore, in Section 2 we present important aspects of our six month study of University College London (UCL) production Grid cluster and their implication in cluster resource management and scheduling. Section 3 details the proposed probabilistic Grid scheduling model, while Section 4 gives preliminary simulation results. The direction of our future work and conclusions are outlined in Section 5, while Section 6 gives references.

## 2. Study of e-Science Grid Application

To establish the feasibility of the statistical prediction methods, and to gain further understanding into the dynamics of job scheduling on a production Grid, a study of workload generated by twenty e-Science projects on a two hundred CPU Sun Grid Engine [9] cluster was undertaken. The fully analysed data includes around 50,000 jobs from the six month period July – Dec 2004, while validation of observations was run on the data up to and including May 2005. Considering that the ultimate source of the computational load was human workflow, a significant degree of correlation, patterns and trends was expected. Our aim was to establish whether these really do occur, and if so whether they can be exploited for predictions with usable accuracy in the context of resource management and scheduling.

In our previous publication [10], we have reported on the data collection and analysis process. The observations here presented are based on the analysis of wall-clock execution times – real time value of process execution time which is larger or, for a perfectly optimised process, equal to the actual CPU time. The data has been examined in two ways: by treating every job consecutively (Figure 1) and by clustering them according to their meta-data (Figure 2). The latter shows clustering based on one of the readily available fields in the accounting file – the Unix group name of the submitter. The group is assigned by the administrator and is loosely related to the e-Science project the users are involved with.

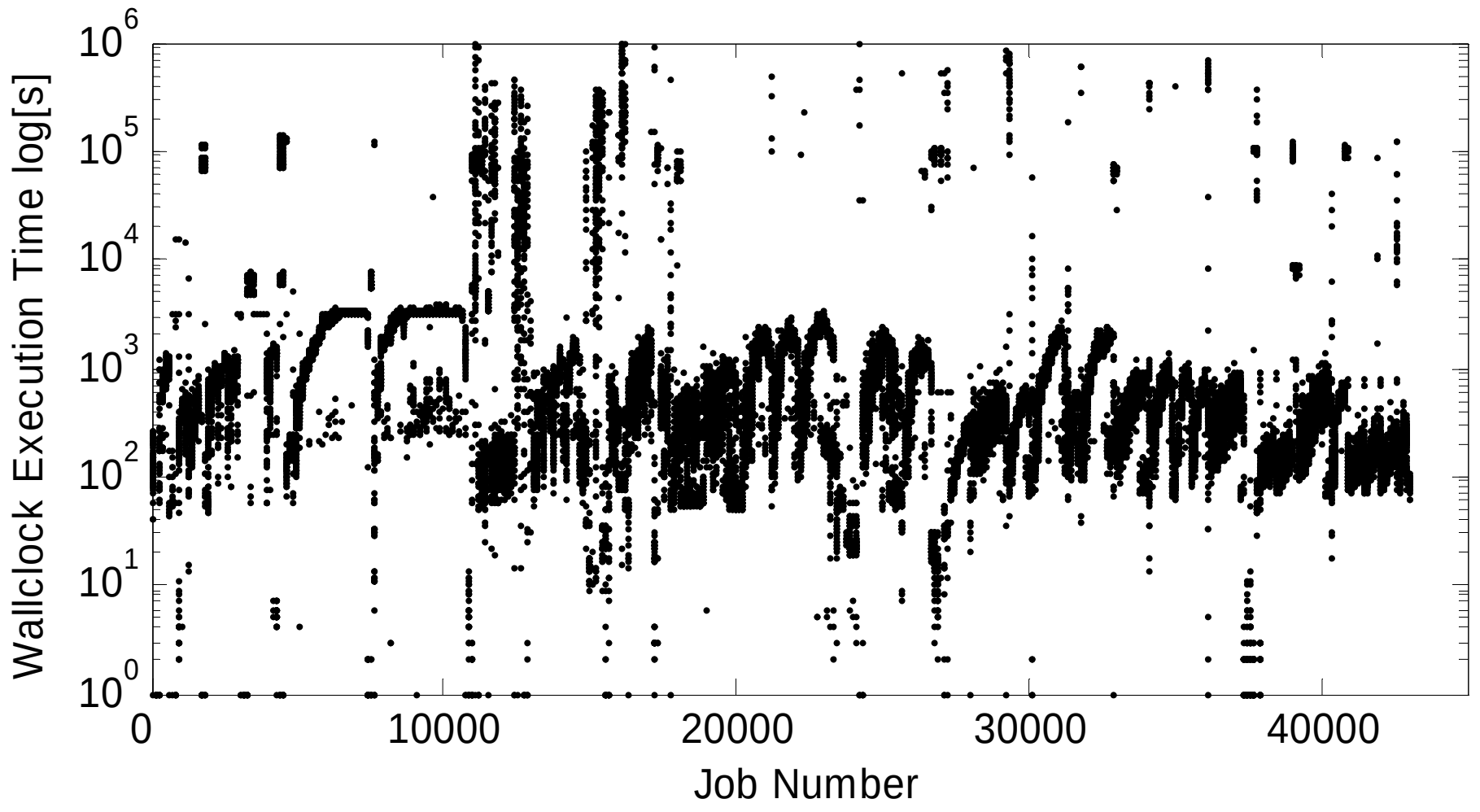

**Fig. 1.** Time-series plot of wall-clock execution times for a six month period

Figure 1 shows a log-normal plot of all recorded job execution times in the observed six month period (43,100 jobs). The first and most immediate feature is the large difference in execution times between the jobs. A small number of submitted jobs either fail, or are test runs of less than 10 seconds (around 4%), while another 2.5% are extremely long jobs running for more than $10^5$ seconds. Overall, execution times span six orders of magnitude and no meaningful model can be developed by treating the data indiscriminately.

Clustering the jobs based on the (group) identity of the submitter results in a clearer view in which differences between execution patterns and job lengths of different groups become obvious (Figure 2). This is but one example of the possible clustering metric used in our scheduler, which has yielded good results for three groups but has not managed to transform the *ocotir* group into a more predictable dataset. We defer further discussion of clustering metrics and methods until Section 4.

Separating raw data based on associated meta-data makes it more suitable for statistical analysis and modelling. In the scheduling context the resulting set contains two distinct modes. In one, execution times are suitable for modelling and prediction by the virtue of either being correlated and randomly distributed within a narrow distribution, or by following a distribution which can be successfully modelled using one of the algorithms implemented in the scheduler. The other mode represents the opposite situation in which abrupt changes in execution times or seemingly random fluctuations prevent predictions within required confidence interval.

As seen in Figure 2, majority of execution points fall in between abrupt discontinuities, and represent an execution instance in which a certain scientific experiment is taking place. The periods of such activity are of most interest to the scheduler, and predictability of behaviour in these regions is crucial. Most encouragingly, these run times show significant levels of short term autocorrelation.

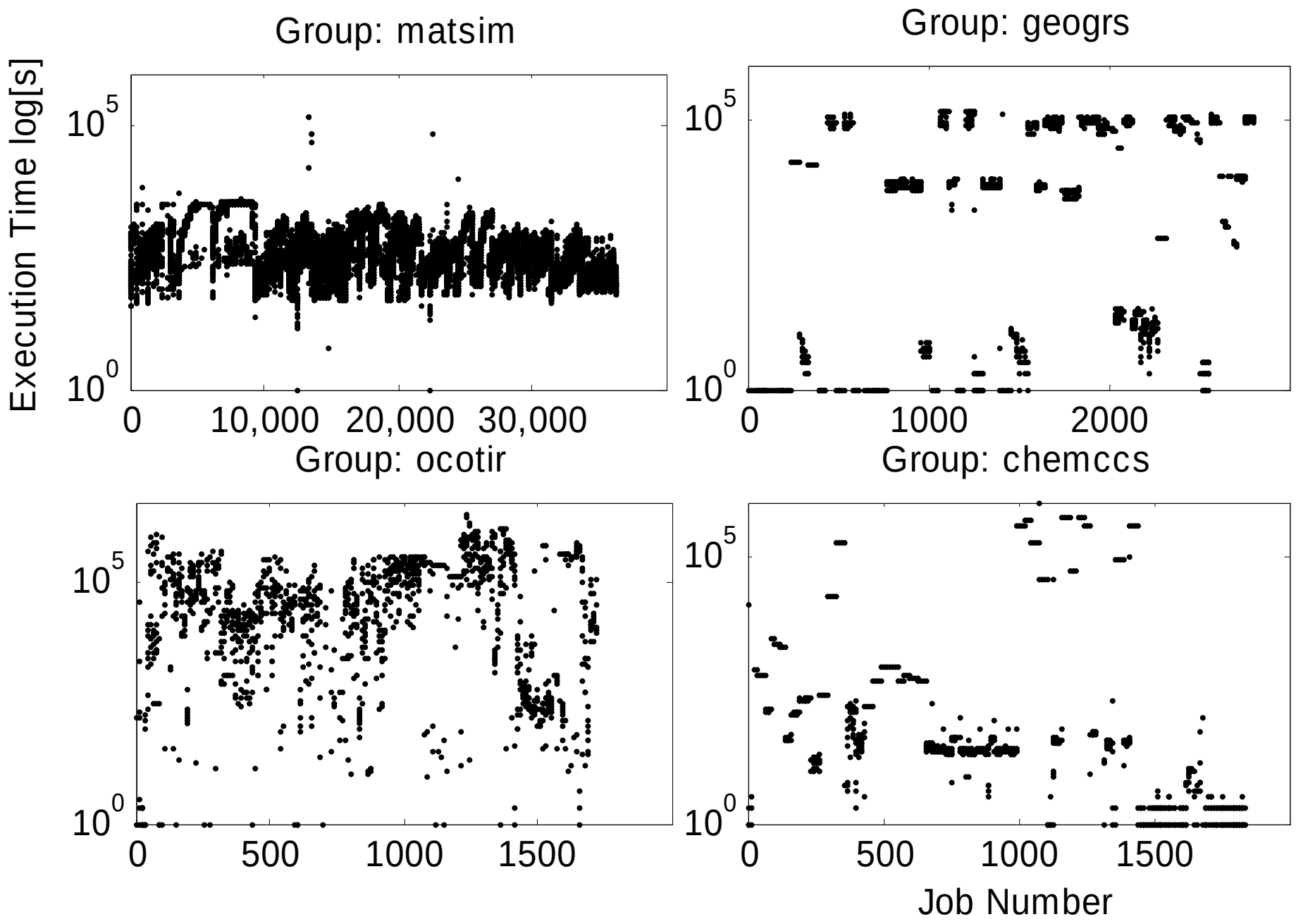

**Fig. 2.** Wall-clock execution times by submitter's credentials

A suitable forecasting algorithm should therefore be able to track and predict future execution times and provide a confidence level for such predictions.

The execution times also exhibit areas with discontinuities and anomalous changes at different frequencies and time scales. Sustained and significant changes in execution times seem to indicate an alteration of the underlying scientific workflow: a new data set might have been introduced, a new scientific objective identified, application updated, or perhaps the group members have taken on a completely different project. This is evident in the group plots in Figure 2 where well differentiated mode changes can be observed. In the scheduling context, these events present instances at which appropriate action should be taken. Conversely, localised and transient deviations from the current statistical distribution of the given process indicate an anomalous event which may be related to an application or resource crash, peculiarity of the data set or other non-permanent behaviour change. One such possible event would be at around job number 350 in the *matsim* group plot in Figure 5. Although these events occur as part of the normal operation of the system, unless their occurrence and nature can be modelled, no immediate action by the scheduler is warranted or justified.

Diversity of jobs in terms of resource requirements and job duration is a prerequisite if a fully probabilistic scheduling approach is to work. In parallel to dimensioning techniques used for telephone networks, statistical multiplexing is possible only if a range of user behaviours results in a variation of service requests in space and time. The cumulative execution time distribution function for all jobs in the six monthly period, as shown in Figure 3, indicates that even a small number of different scientific projects and applications give rise to a heterogeneous environment in terms of the job length and complexity. The distribution has a very good linear fit from 10% to 90% of likely observations, covering execution times from around 50 seconds to 5000 seconds. This implies any execution time duration from that range is as likely to occur as any other and ensures that the scheduler will work on a broad mix of job sizes suitable for out-of-order execution and backfilling.

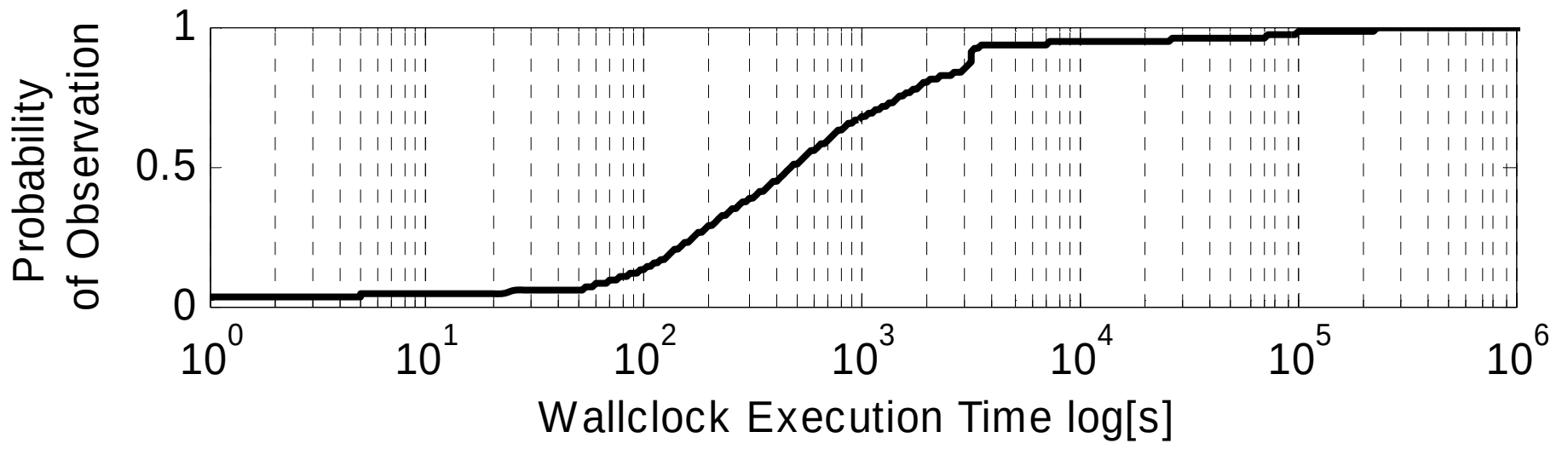

**Fig. 3.** Job wall-clock execution times cumulative distribution function

The linearity of the cumulative distribution function is artificially altered at the top and bottom ends by the user's own limiting of the range of applications they can sensibly run on this Grid – it is unlikely they will invest the effort of setting up and running very small jobs which can equally well be run on their own workstations, while at the same time abstaining from running jobs which on this hardware would take impractically long time to complete. We anticipate that the penetration of interoperable Grid middleware into ever lower as well as higher end platforms will enable even broader range of applications and extend the linearity range, but that user's local decision making will invariably impose tail-off effects which need to be suitably handled.

All aforementioned properties of the data were present throughout the six monthly period of the job trace. However, events such as changes in the mode of operation and the frequencies of anomalous, out-of-distribution execution times showed significant temporal locality. Some of the causes for clustering of these events will be further discussed in Section 4, but such behaviour has to be considered when building a probabilistic scheduler model. Similarly, classes of applications, users or groups evolve their behaviour altering the parameters of the best fit distribution of their execution times, as well as shifting them from one statistical distribution to another. As a consequence, rather than applying fixed assignments, the scheduling model must be able to adapt and evolve its model of any given class of jobs through time and according to the current state of the Grid environment.

Overall, this study of e-Science Grid applications, their execution times and associated workflow has shown the repetitive nature of jobs and their correlation with historical runs as well as related meta-data. Following section will present a suitable scheduling model able to use this data to predict the execution time of a newly submitted job in order to maximise the likelihood of successful completion before the requested deadline.

## 3. Probabilistic Scheduling Model

Our motivation for developing deadline based scheduling was to extend natural human workflow into the Grid environment. Scientific experiments, and business projects especially, run to certain deadlines and monetary constraints. While overall completion of a complex task may be due in several years, sustained progress may be dependant on a set of smaller tasks finishing by the end of the lunch hour, by next morning or over the weekend. As Grid systems are usually space- and time-shared environments, prioritising jobs is important aspect of current batch schedulers. These policies, however granular and flexible they may be, assign fixed weights to certain user groups and do not automatically account for changing priorities of such groups on a task to task basis. The deadline implies a priority measure for each job, and should prevent a highly urgent job from a low priority user group being blocked by a non-critical job submitted by a high priority user. To prevent the user "playing" the scheduler, we envisage that an economy system will complement the deadline scheduler by assigning a nominal currency cost to computational resources and execution priorities. Although this is an important line of research in Grid scheduling [11], it is not within the scope of our work.

Our probabilistic scheduling model, shown in Figure 4, embraces the stochastic nature of the underlying execution time data, anomaly thresholds and job clustering to offer a probabilistic guarantee of deadline adherence: a job will be allocated enough time and resources to finish by a certain deadline based on its historical performance and the confidence level depending on the priority (or economy constraints) of the user. The methodology tries to extract maximum value from the data already routinely collected in Grid clusters. Job meta-data kept in the accounting files is actively analysed for patterns and correlations, while schedule performance and monitoring data is fed back from the Grid management layer into the scheduler forming a control loop influencing scheduling decisions.

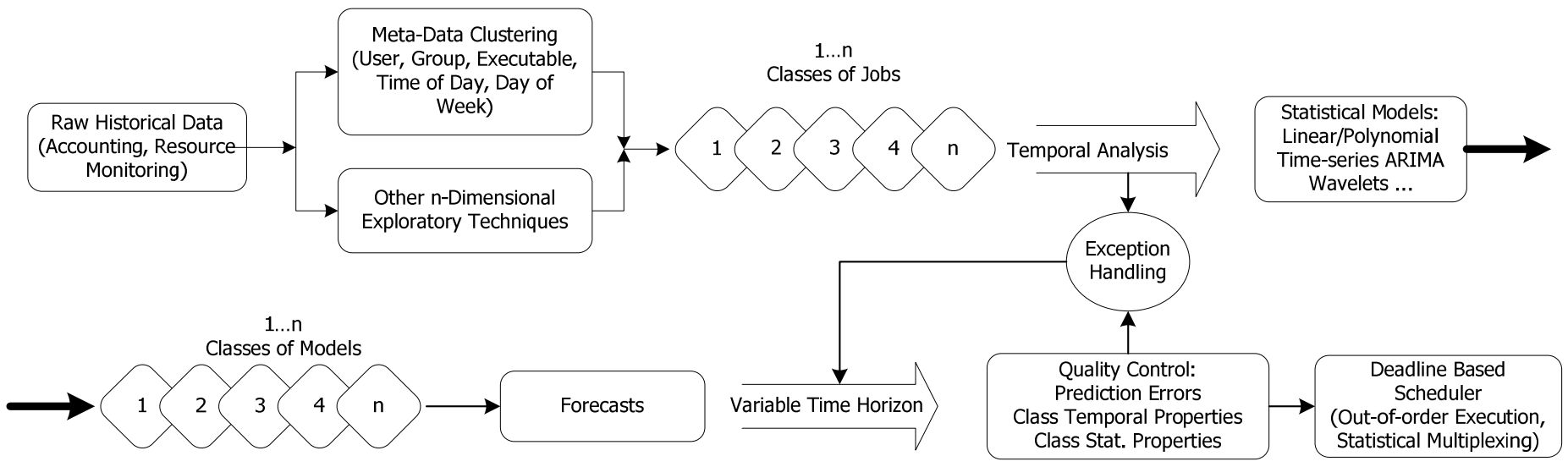

**Fig. 4.** Probabilistic grid scheduling architecture diagram

The model is envisaged as a core component of a stand-alone scheduler, or integrated into a modular scheduler like ICENI [12]. It consists of three main functional parts: job classification & clustering, forecasting & predictions, and anomaly detection.

### 3.1. Job Classification and Clustering

Job classification and clustering module is responsible for locating correlations between job execution times, resource utilisation and its associated meta-data. Meta-data is extracted from the accounting file, and while it will differ slightly between Grid middleware it will contain at least the job submitter credentials, submitting time stamp and the submitting host. The result of this clustering step is a grouping of jobs with similar statistical properties, arrival rates and execution time distributions, and association of this class of jobs with some meta-data criteria.

One example would be that at the time of invocation of the algorithm, a certain user is submitting queries to a protein database from her workstation and has been doing so repeatedly over the period of several work days. Any subsequent job submission from her username and her workstation, invoking such application within normal workday office hours will initially be considered as part of the same workflow and likely to behave in a way consistent with the developed model. A number of orthogonal models can be developed, and an alternative one may be describing a more intensive job set this user is submitting overnight or over the weekend.

Currently we have developed raw data import filters for Ganglia Cluster Monitoring system and Sun Grid Engine, the middleware of choice at the UCL Grid installation, but other open data sources could as easily be accommodated.

### 3.2. Prediction Models and Forecasts

The forecasting module produces a prediction, with a defined confidence level, of the job's execution time and resource utilisation on a known hardware node based on historical data. For each class of applications identified by the clustering algorithm, a number of different models are fitted. Considering that our context is on-line, fast and automated forecasting, we have opted for univariate prediction models that can be parameterised using automated search or heuristic methods. We have tested several algorithms which should suite a wide range of application behaviour.

A range of simple and fast methods was used, these include mean or median estimation technique based on different order polynomials and exponential smoothing with or without trends (Holt method) [13]. The number of required parameters is small, and their estimation a quick iterative procedure. The performance of these algorithms will be discussed in more detail in Section 4, but despite their simplicity they yield acceptable results.

A more complex prediction method treats measures of historic execution times as time-series values and builds a model based on auto-regressive (AR) and moving-average (MA) approaches. These have been used extensively in modelling of stochastic processes (from financial data to seismic observations), and have the ability of estimating variations of a value over a long prediction horizon. The parameterisation of AR and MA processes is computationally expensive, but heuristic approaches can be used to optimise the search and develop a fully automated system. We have successfully used combined AR, MA, ARMA and ARIMA models as further described in Box and Jenkins [14].

As a future development, we intend to investigate the use of wavelets in modelling and forecasting execution times and resource requirements. With faster hardware and more efficient algorithms, wavelet's ability to offer superior fit across the frequency spectrum is becoming more appealing despite their computational complexity.

Given the variations observed in the execution time data, it is not likely that any one model can be pre-selected for any given class of jobs or that any one job class will remain well described by a particular model for a prolonged period of time. To accommodate this dynamic nature of Grid applications, all forecasting methods will work together in an expert system, giving simultaneous predictions with one or more differently parameterised models. A feedback loop from the monitoring system will report on the actual values observed once the job has run, and these will be used to positively enforce the best algorithm and model parameterisation.

### 3.3. Mode Changes and Anomaly Detection

The analysis of Grid application in Section 2 has shown that execution times and resource utilisation of submitted jobs can vary significantly in short periods of time. The reasons for such behaviour are many, but in a truly probabilistic approach these events should be seen as a less likely occurrence rather than an anomaly.

In the predictive scheduling context, events of special interest are transitions between different modes of operation within a specific class of jobs. From the large changes in job execution times between different modes of operation for each user group, as shown in Figure 2, it is clear that significant prediction errors will occur while the forecasting algorithm acquires enough training data points.

We define these transients as areas in which a trained forecasting model shows sudden, significant and sustained drop in the accuracy of predictions. At this point, it is likely that for whatever underlying reason the job class has shifted significantly and is not adequately described by the model any longer.

The primary difficulty is in establishing whether an event is a short-lived deviation from the current model or a significant change of behaviour is occurring. Immediate indication is given by the increase in forecasting error, and this can be further corroborated by analysing job meta-data for changes (executable name or command line parameters for example). We have expanded this approach with the use of smoothing or moving average functions which acts as a low-pass filter on the execution times series. By monitoring the difference between the actual data points and values produced by a smoothing or moving average function a trigger can be fired when a significant deviation is observed. Black diamond markers in Figure 5 indicate positives from the anomaly detection module. This is by no means the only or the most sophisticated approach to detecting sudden changes, and a method based on wavelet analysis is being pursued as further work on this topic.

Regardless of the detection method used, the handling of anomalous events is guided by the principle that positive action should only be taken if it would lead to an improvement in the accuracy of predictions. On detecting a sudden and significant forecast error, the system will immediately start training an alternative model and produce forecasts in parallel. If the event turns out to be a short lived deviation, the new model will not have enough training points and will not be able to produce results better that the model already in use. Should the event be a mode change, the old model will be phased out at the point at which the new one has started producing superior forecasts. Together with this staged discarding of training data, anomaly detection module reacts to the reduced confidence in forecast by shrinking the prediction horizon and communicating the drop in the level of certainty to the scheduler which can ensure quality of service by applying a larger safety margin to forecasted values and a more conservative schedule.

## 4. Prediction Algorithm Simulation Results

The functionality so far described has been implemented as an emulation platform using MathWorks MATLAB™ and add-in components programmed in C. Testing and improvement of the model is done on a subset of data points in which particular behaviour or events of interest, such as mode changes, sudden deviations from modelled behaviour or execution times with different distributions, occur. The data collected from the UCL Grid cluster is fed into the simulation via a trace reply system, and resulting predictions compared with the actual observed values.

The functionality so far developed includes:

- Data acquisition from the Sun Grid Engine accounting files and Ganglia round-robin databases [15]
- Meta-data grouping according to submitting username/group
- Forecasting using third order polynomial and automated ARMA algorithm
- Anomaly and mode change detection using prediction error tracking and Lowess smoothing [16]

The quality of the prediction algorithm and its parameterisation is assessed using a summary plot produced by the simulation, shown in Figures 5 and 6. It shows one-step-ahead predicted and actual execution times, percentage prediction errors, error values histogram and data points flagged as anomalous values. In Figure 5, the forecasting method used is third order polynomial fitted to 10 historical data points and anomaly detection was based on fixed threshold values of 50% for prediction errors and 25% for deviation from a Lowess smoothed curve. The performance of the prediction algorithm was in the range of 25% to 75% error rate except in the areas in which anomalous behaviour or mode change occurs. However, the variance of the actual values is exaggerated by the piece-wise polynomial trend fitting resulting in a considerably noisy forecasted series. The anomaly detection algorithm has given good results, indicating the point at which mode change occurs and points at which the model has returned unsatisfactory predictions. A number of false positives are also present, but their effect is mitigated by the policy of staged transition to the new forecasting model. As part of our further work we will investigate a more sophisticated anomaly triggers with floating thresholds.

A similar summary plot in Figure 6 demonstrates the performance of a 4th order autoregressive, 4th order moving average prediction model. The model produces very good results with consistently small prediction errors, fast convergence after mode changes and smaller variance of the prediction errors, as indicated by the error value histogram. The order of the model was selected manually in this case, although we are testing different heuristic methods to automate this process. There is an increased computational cost associated with a more complex model such as ARMA, but we defer a more thorough investigation of its effects on the overall scheduling model until later in the optimisation phase.

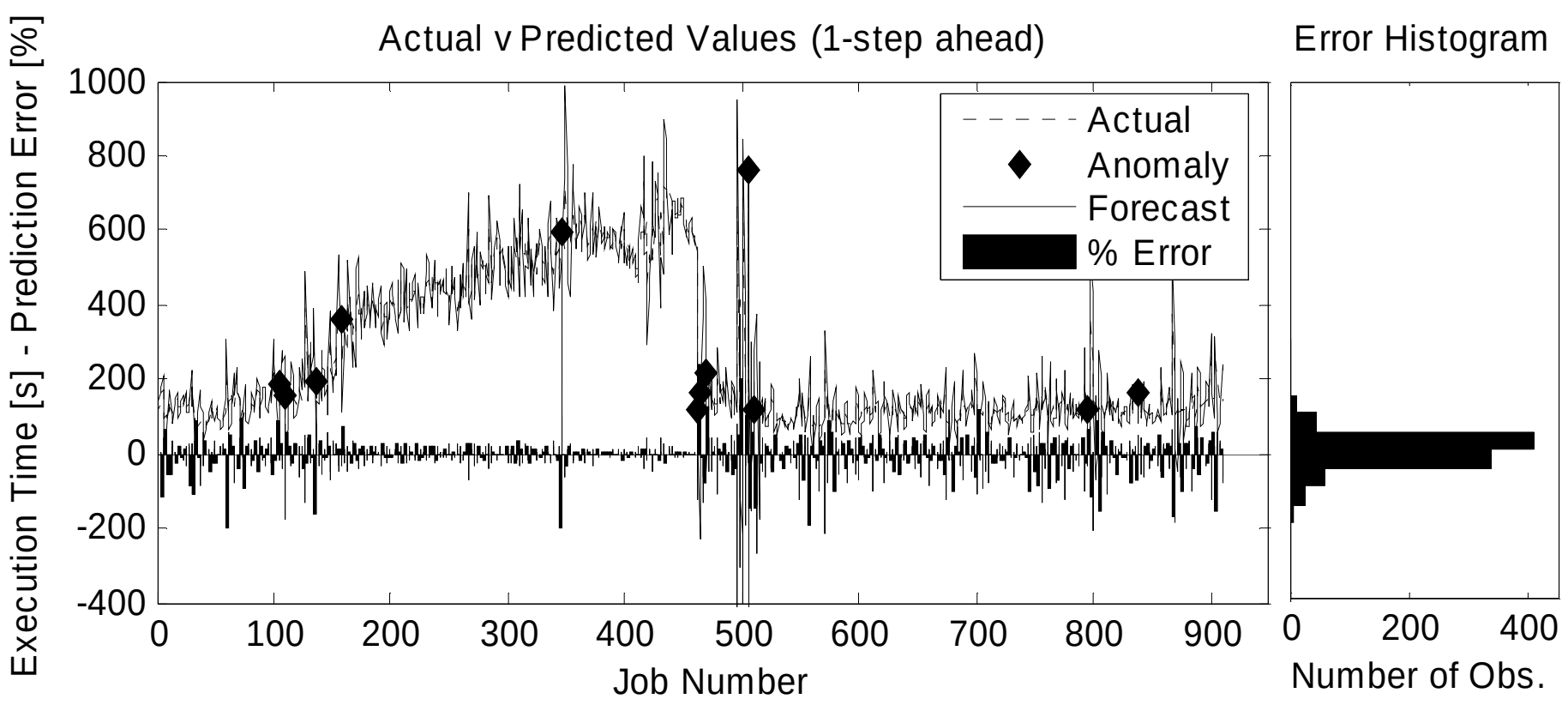

**Fig. 5.** 3rd order polynomial forecasting performance plot

The purpose of these preliminary tests was to provide us with a guide as to the predictability of the execution times and the areas in which largest improvements can be made. No extensive optimisation of the model parameters was done, and for reasons of implementation simplicity anomaly detection was implemented purely as an observation function and did not trigger any actions. Over a large number of samples the prediction algorithms achieved an almost zero mean percentage error rate indicating that in a normal deployment scenario with a constant stream of requests underestimates on certain jobs will cancel overestimates on others. This further supports our probabilistic policy which guarantees that completion within a deadline will follow a certain distribution function, but does not offer assurance that any one specific job will be executed on time.

Development of the simulation raised several practical issues with the data and algorithms involved. Due to the lack of standardization in the evolving Grid community, utilisation and accounting records are not uniform, complete or fully documented. This prevents the exchange of workload traces across different clusters and verification of developed model on alternative Grid installations. In depth analysis of accounting logs revealed significant opacity in the executable identification and run time parameters. In the case of Sun Grid Engine, a single field maintains the name of the executable submitted, which is usually a generic shell script, thus hampering the effort to identify changes in the name of the application run, possible recompilation of the application or a change in run time parameters. Sharing of usernames and running of jobs under generic credentials also seems to be a widely used practice, and has an adverse effect on the granularity of our job classification and clustering module. The forecasting algorithms tested handled different job distributions well. The simplicity and speed of trend estimation makes it an appealing option for classes of numerous short processes where the delay in creating a more complex model may not be justified compared to the gain in accuracy and the total execution time of the process. However, these simpler methods are particularly susceptible to the ringing effect from large deviations in the data.

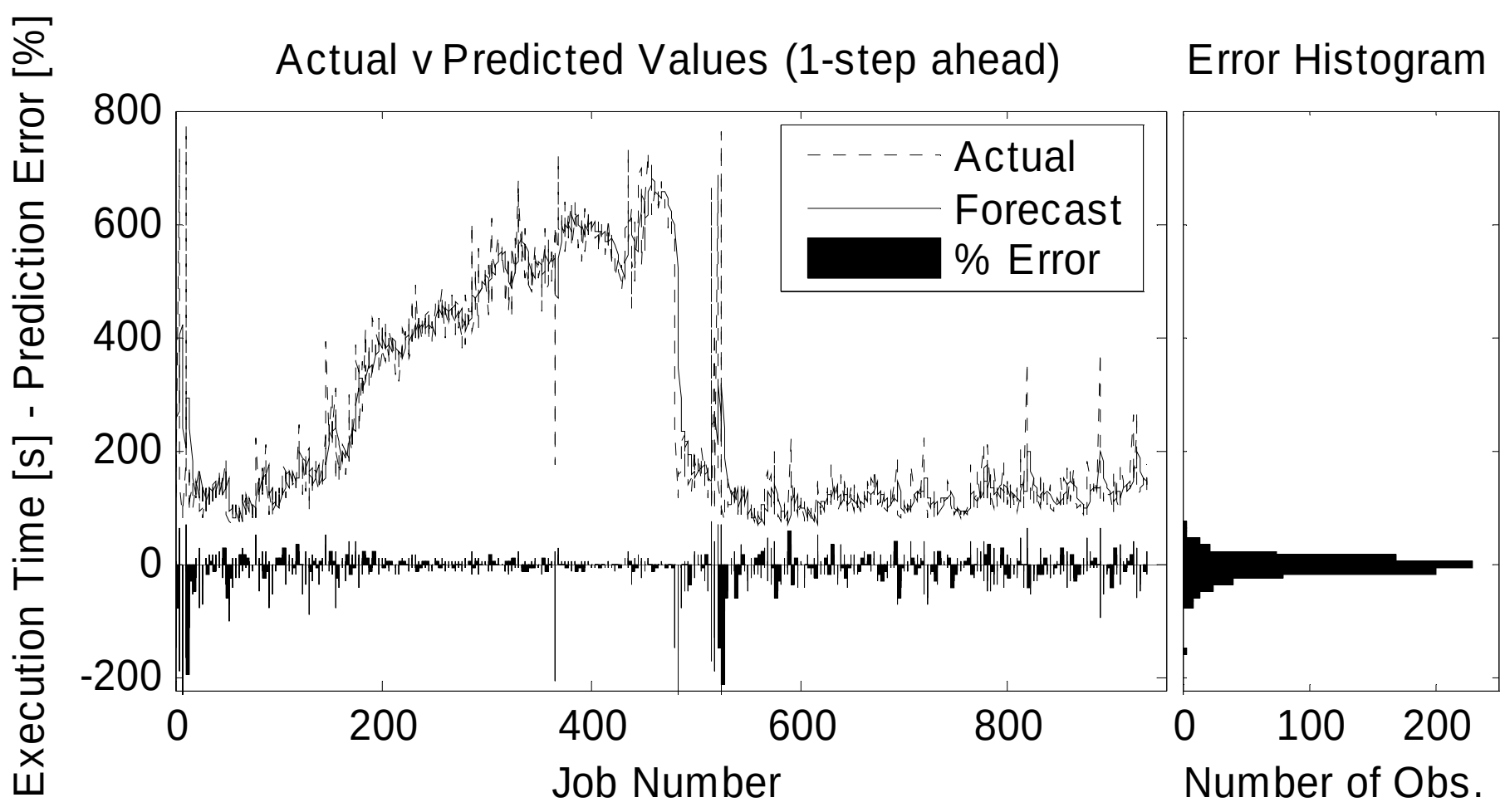

**Fig. 6.** 4th order autoregressive, 4th order moving average forecasting performance plot

## 5. Issues and Conclusions

Our research group has previously looked at the issues of resource discovery and intrusion detection, proposing distributed and self-organising approaches to these problems [20, 21]. Presented scheduling model has been motivated by the need for a light-weight, autonomous and self-managing scheduler suitable for the dynamic nature of service-orientated utility Grid environments and their numerous heterogeneous applications.

To understand the demands that such Grid may experience, we have studied the computational load on a multi-purpose production Grid serving a number of different e-Science projects. We consider the data sample to be both exhaustive and relevant; in addition to the six months job set directly used in this paper our ongoing analysis confirms observed patterns and features of the data. The Grid cluster in question is a multi-purpose facility, connected to the UK's national Grid infrastructure, and as such is used by collaborating projects throughout Europe. The authors have approached several other multi-purpose Grid operators for utilisation data, and preliminary analysis from one of the largest pan-European Grid and one of the earliest adopters of Grid technology in the banking sector indicate very similar job characteristics as described in this paper.

This study has established that the heterogeneous overall job population can be separated into smaller classes with more consistent statistical properties by using meta-data relating to submitted jobs. Cumulative distribution plots also indicate that job execution times may have scale-free properties. By using traces of e-Science applications run on the production Grid in our probabilistic scheduler emulation environment, we have demonstrated that a statistical model can be used to describe the properties of these job classes and deliver predictions on the execution times of jobs pending in the scheduling queue. This will form the basis of our future work on the deadline/economy based scheduler.

The scheduling approach presented is not dependant or based on any specific Grid middleware or scheduler implementation. It is rather a methodology for converting accounting and monitoring information normally collected into scheduling intelligence able to support self-managing deadline scheduling. Depending on the operating environment, the implementation framework can be modified to use as much meta-data as iis available, or as much local scheduling capability as the system will support (such as reservations, checkpointing or job migration). The approach presented departs from current practice of scheduling on the Grid by treating the workload in a probabilistic manner similar to other networked service delivery platforms. The system is able to offer added value to the end user in the form of job deadline execution in an autonomous and self-managed way, while providing the Grid operator with the ability to satisfy different service level requirements at different "cost" points.

The feasibility and value of our approach has been strengthened by encouraging results reported in this paper. We remain confident that given the vertical expansion of the utility computing approaches across hardware of different computational power, and associated broadening of the applications spectrum, statistical treatment of application load will find its use in a range of resource management tasks from planning and provisioning at longer time scales to scheduling at cluster, host and CPU levels at increasingly finer granularity.

Although we have not directly treated the economy part of the scheduling problem, it is nevertheless an important aspect required for delivering deadline scheduling. Without a notion of nominal cost, each user could request their jobs to be turned around as quickly as possible, in which case the whole approach would revert back to a batch scheduler.

The means of assessing the feasibility of the deadline is an equally important question and an area of our future work. Due to its intended target environment, probabilistic nature and inability to guarantee adherence to the deadline for any particular and singular job, our approach may not be suitable for mission-critical or real-time systems, or for execution on very specialised hardware for which (semi-)manual job admission is used.

The scheduling implementation described in this paper would act as a local level scheduler on a Grid connected cluster able to develop statistical models of applications running in that specific, homogeneous, environment. This is in line with current Grid provisioning practices whereby these clusters are confederated into larger heterogeneous Grids. We envisage an extension of our system to provide higher level scheduling capability by enabling local schedulers to communicate their best deadlines and nominative prices as a bid value to an economy based super-scheduler. We are working on a suitable encoding of the statistical model describing a certain class of jobs and its effective communication to other local level scheduler using one of the distributed protocols developed by our research group [20, 21].

Ultimately, regardless of the potential for data clustering based on orthogonal meta-data, certain number of jobs for which no suitable model can be developed will remain. These can be dealt with in a number of ways, from running on dedicated nodes to hard-limiting, but at the very least the scheduler will be able to recognise on job submission (within given confidence level) that a certain process belongs to such group and hence take appropriate actions to minimise the impact of its unpredictability.

## 6. Related and Future Work

Previous work in the area of Grid resource management has recognised the need for a departure from first-in-first-out batch approach, and has established the benefit of predictive scheduling approaches. AppLeS [7,17] relies on custom developed models of target application and hardware platform to deliver scheduling to a target turnaround time, but requires extensive modification of application's source code. Nimrod/G [18, 19] uses a novel "budget" metric to support a market-based framework for deadline scheduling. However, its scope is limited to parametric study applications whose execution times are very narrowly distributed, and generally independent of the input parameters. ICENI scheduler [12] provides a general framework for a predictive scheduler and implements a forecasting module based on directed acyclic graph analysis of the running application. Running time advisory (RTA) [22] uses similar time-series techniques as presented here but is focused on estimating task execution time as a function of current and forecasted host load. RTA predictions are computed from a nominal execution time of a task on an unloaded machine, but although this value is essential to the algorithm, RTA authors propose no method of calculating it. Our work produces exactly those estimates and can be extended with RTA-like approaches in cases where Grid nodes are non-dedicated hosts.

Deployment on a production system with enough jobs with varying characteristics at such an early stage in concept development may not be suitable or feasible. Having established the methods for modelling of historical job execution times, our future work will be centred on developing a validation tool based on real job traces from production environments. We would like to further investigate cyclic effects of job submissions, and will investigate ways of circumventing opacity and non-standardisation in current Grid middleware. In this aspect, global identification of job submitter through certificate's Distinguished Names will be looked at. A complementary line of research will look at deeper workflow analysis and possible interaction with Grid workflow tools in order to obtain better understanding of user's actions and anticipation of next steps he is likely to take. Finally, through a deeper analysis of statistics of anomalous events, a heuristic model for setting appropriate trigger thresholds will be investigated.